\documentclass{jpp}
\usepackage{graphicx}
\usepackage{epstopdf, epsfig}
\usepackage{hyperref}

\shorttitle{Standing autoresonant plasma waves}
\shortauthor{L. Friedland and A. G. Shagalov}

\title{Standing autoresonant plasma waves}

\author{L. Friedland\aff{1}
  \corresp{\email{lazar@mail.huji.ac.il}},
  A. G. Shagalov\aff{2,3}}

\affiliation{\aff{1}Racah Institute of Physics, Hebrew University of Jerusalem,
Jerusalem 91904, Israel
\aff{2}Institute of Metal Physics, Ekaterinburg 620990, Russian Federation
\aff{3}Ural Federal University, Mira 19, Ekaterinburg 620002, Russian Federation}

\begin{document}

\maketitle

\begin{abstract}
Formation and control of strongly nonlinear standing plasma waves (SPWs)
from a trivial equilibrium by a chirped frequency drive is discussed. If the
drive amplitude exceeds a threshold, after passage through the linear
resonance in this system, the excited wave preserves the phase-locking with
the drive yielding a controlled growth of the wave amplitude. We illustrate
these autoresonant waves via Vlasov-Poisson simulations showing formation of
sharply peaked excitations with local electron density maxima significantly
exceeding the unperturbed plasma density. The Whitham's averaged variational
approach applied to a simplified water bag model yields the weakly nonlinear
evolution of the autoresonant SPWs and the autoresonance threshold. If the
chirped driving frequency approaches some constant level, the driven SPW
saturates at a target amplitude avoiding the kinetic wave breaking.
\end{abstract}

\section{Introduction}
Plasmas can sustain laser intensities many orders of magnitude larger than
typical solid state optical components. This makes plasmas attractive for
manipulation and control of intense laser beams, but requires formation of
large and stable electron density structures in the plasma to impact the
propagation of the laser light significantly. A typical approach to this
problem is via ponderomotive forces using additional driving laser beams.
Several applications of \textit{plasma photonics} based on light scattering
off electron density structures have been proposed including short pulse
amplification via the resonant excitation of plasma waves \citep{Malkin},
transient plasma gratings \citep{Lehmann2016}, crossed-beam energy transfer
for symmetry control in ICF \citep{Michel2009}, and, more recently,
plasma-based polarization control %
\citep{Michel2014,Turnbull2016,Turnbull2017,Lehmann2018}. The efficiency is
an important factor in all these applications, as the goal is to create the
largest amplitude plasma density perturbation using smallest possible driver
intensities. Recently, it was shown that very large amplitude travelling or
standing ion acoustic waves (SIAW) can be excited via the autoresonance (AR)
approach \citep{Lazar142,Lazar151,Lazar158}. The latter exploits the salient
property of nonlinear waves to stay in resonance with driving perturbations
if some parameter in the system (e.g., the driving frequency) varies in time
[for a review of several AR applications see \citep{Lazar92,Lazar125}]. Such
systems preserve the resonance with the drive via self-adjustment of the
driven wave amplitude leading to a large amplitude amplification even when
the driver is relatively weak. In this paper, we will discuss the AR
approach to excitation of very large amplitude standing plasma waves (SPW)
using both the kinetic and a simplified water bag models. The nonlinear
SIAWs \citep{Lazar158} and SPWs [see a related analysis of nonlinear
Trivelpiece-Gould waves \citep{Dubin}] are much more complex than their
traveling wave counterparts. For example, the autoresonant SIAW driven to
large amplitudes by a weak, chirped frequency standing ponderomotive wave is
a two-phase solution, each phase locked to one of the traveling waves
comprising the drive \citep{Lazar158}. The peak electron density in these
waves can reach several times the initial plasma density. We will show that
similar giant plasma structures can be excited in the form of autoresonant
SPWs.

The paper is organized as follows. In Sec. 2, we will illustrate
autoresonant SPWs in Vlasov-Poisson simulations and compare the results to
those from the a simplified water bag model presented in Sec. 3. We will
discuss weakly nonlinear SPWs in Sec. 4 and use their form as an ansatz in
developing a weakly nonlinear theory of driven-chirped SPWs based on the
Whitham's averaged variational principle \citep{Whitham} in Sec. 5. In Sec.
6, we reduce a system of slow coupled amplitude-phase mismatch equations
yielding the autoresonance threshold on the driving amplitude for excitation
of autoresonant SPWs. In the same section, we will also discuss the control
of large amplitude autoresonant SPW by tailoring the time dependence of the
driving frequency. Finally, Sec. 7 will present our conclusions.

\section{Standing autoresonant plasma waves in Vlasov-Poison simulations}

Our study of autoresonant SPWs is motivated by numerical simulations of the
following one-dimensional Vlasov-Poisson (VP) system describing an
externally driven plasma wave%
\begin{equation}
f_{t}+uf_{x}+(\varphi +\varphi _{d})_{x}f_{u}=0,\varphi _{xx}=\kappa
^{2}\varphi +\int fdu-1.  \label{VP}
\end{equation}%
\begin{figure}
\includegraphics[width=9cm]{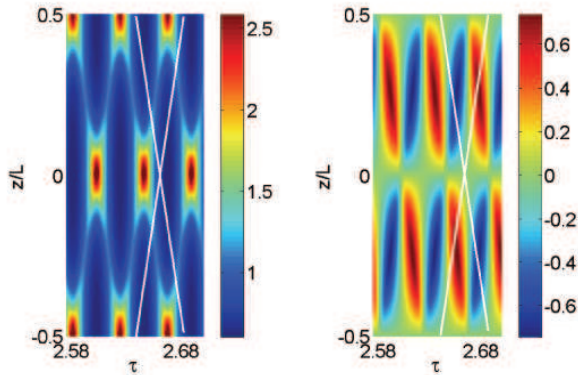}
\caption{(color online) The colormap of the waveforms of the electron
density $n$ (left panel) and fluid velocity $v$ (right panel) in the final
time interval $\protect\delta \protect\tau =0.125$ of the simulation. The
slopes of the white lines are the phase velocities of the two traveling
waves comprising the drive.}
\end{figure}
\begin{figure}
\includegraphics[width=9cm]{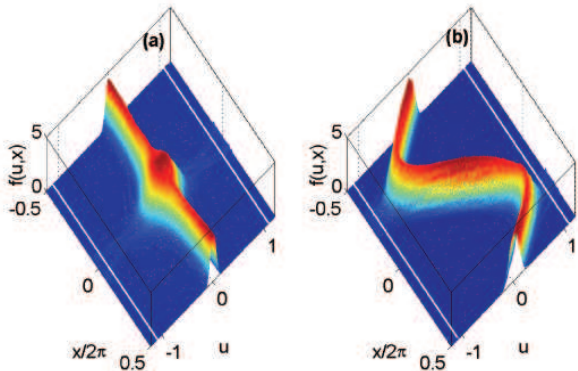}
\caption{(color online) The snapshots of the electron distribution function
at two times $\protect\tau =2.69$ [panel (a)] and $2.7$ [panel (b)] during
the last period of the wave oscillation in Fig. 1. The fluid velocity
vanishes in panel (a), while the electron density reaches its maximal value
of $2.5$. In contrast, in panel (b), the fluid velocity $|v|$ is at its
maximum at two locations in $x$. The straight white lines are at the locations
of the phase velocities $\pm \protect\omega _{d}/k$}
\end{figure}
Here we assume constant ion density, $f$ and $\varphi $ are the electron
velocity distribution and the electric potential, and $\varphi
_{d}=2\varepsilon \cos (kx)\cos \theta _{d}$, where $\theta _{d}=kx-\int
\omega _{d}dt$ is a small amplitude standing wave-like ponderomotive
potential having a slowly varying frequency $\omega _{d}(t)$. All dependent
and independent variables in (\ref{VP}) are dimensionless, such that the
position, time, and velocity are rescaled with respect to $1/k$, the inverse
plasma frequency $\omega _{p}^{-1}=\sqrt{m/(4\pi e^{2}n_{0})}$ ($n_{0}$
being the ion density), and $\omega _{p}/k$. Then, in the dimensionless form
of the driving phase, $k=1$ and $\omega _{d}$ is rescaled by $\omega _{p}$.
The distribution function and the potentials in (\ref{VP}) are rescaled with
respect to $kn_{0}/\omega _{p}$, and $m\omega _{p}^{2}/(ek^{2})$,
respectively. We have also added an effective screening term $\kappa
^{2}\varphi $ in Poisson equation modeling the typical finite radial extent
of the ponderomotive potential \citep{Dubin}. We assume $2\pi $ periodicity
in $x$ and solve the time evolution problem, subject to simple initial
equilibrium $\varphi (x,0)=0$ and $f(u,x,0)=(2\pi \sigma ^{2})^{-1/2}\exp
(-u^{2}/2\sigma ^{2})$, where $\sigma =k\lambda _{D}$ ($\lambda
_{D}=u_{e}/\omega _{p}$ being the Debye length and $u_{e}$ the initial
electron thermal velocity). Note that $\sigma $, $\kappa $ and the driving
parameters fully define our rescaled, dimensionless problem. Finally, we are
interested in the driving frequency of the order of the plasma frequency
and, consequently, assume $\sigma \ll 1$ to avoid the kinetic Landau
resonance ($\omega _{d}/k\approx u$) initially.

We have applied our VP code \citep{Lazar111} for solving this problem
numerically and show some results of the simulations in Figs. 1-4 for $%
\sigma =0.1$, $\kappa =0$ (the effect of the nonzero $\kappa $ will be
discussed later), and driving frequency (dimensionless) $\omega _{d}=\omega
_{0}+\alpha t$, where $\omega _{0}=\sqrt{(1+\kappa ^{2})^{-1}+3\sigma ^{2}}$
is the normalized plasma wave frequency in the linearized warm fluid model
[see Eq. (\ref{4.4})], while $\alpha =0.00005$ and $\varepsilon =0.0021$. We
present the results of the simulations versus slow time $\tau =\alpha ^{1/2}t
$ (a convenient representation in AR problems), as the system evolves
between initial $\tau _{0}=-5$ and final $\tau _{1}=2.7$ times (the total
dimensionless time in this simulation is $\Delta t=7.7/\alpha ^{1/2}=1089$,
i.e., $173$ periods of plasma oscillations). Figure 1 shows the waveform of
the electron density $n=\int fdu$ and fluid velocity $v=\int ufdu$ in a
small time window $\delta \tau =0.125\,$\ of the evolution just before
reaching $\tau _{1}$. One can see a very large amplitude SPW with its peak
electron density reaching $2.5$ times the initial plasma density (unity in
our dimensionless problem). Furthermore, as expected, for a given time, the
solutions are $2\pi $ periodic in $x$. But there also exist two directions
shown by white lines in the figure with slopes $dx/d\tau =\pm \omega
_{d}/\alpha ^{1/2}$ (these slopes are the phase velocities of the two
traveling waves comprising the drive) along which the solutions are $\pi $
periodic. This suggests that $u,n$ are periodic functions of \ three
arguments $\Theta _{1},\Theta _{2},x$, where $\Theta _{1,2}=x\pm \int \omega
_{d}t$ and the solutions are $2\pi $ periodic in $\Theta _{1,2}$ and $\pi $
periodic in $x.$ Similar two-phase solutions, with each phase $\Theta _{1,2}$
locked to one of the phases of the waves comprising the drive were also
observed in autoresonant SIAWs \citep{Lazar158}. The occurrence of
multi-phase solutions is known in the theory of some partial differential
equations \citep{Novikov}.

Additional results of the simulations are presented in Fig. 2, showing the
snapshots of the electron distribution function at two different times
inside $\delta \tau $. Figure 2a corresponds to the time when the fluid
velocity vanishes, while the electron density reaches its maximal value of $%
2.5$, while at the time in Fig. 2b the fluid velocity $|v|$ is at its
maximum at two locations in $x$. The white\ horizontal lines in Fig. 2 show
the locations of the phase velocities $\pm \omega _{d}/k$ associated with
the traveling waves comprising the drive. One can see in Fig. 2b that a
small fraction of the electrons in the tail of the distribution reaches the
Landau resonance, indicating proximity to the kinetic wave breaking.
\begin{figure}
\includegraphics[width=9cm]{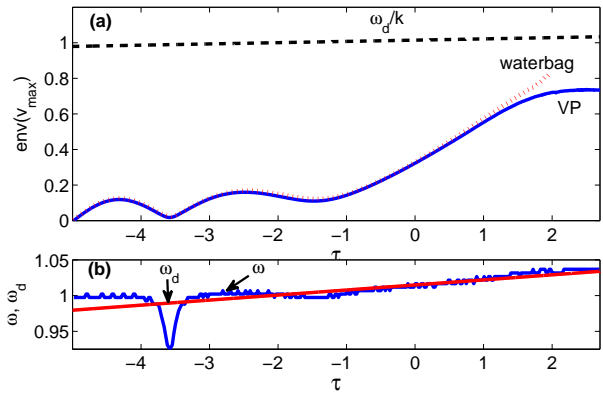}
\caption{(color online) The excitation of the autoresonant SPW for driving
amplitude above the AR threshold. Panel (a): the envelope of the maxima of
the electron fluid velocity $v$ versus slow time $\protect\tau =\protect%
\alpha ^{1/2}t$ in VP simulations (blue solid line) are compared to water bag
model simulations (dotted red line). The dashed line represents the phase
velocity $\protect\omega _{d}/k$ of the driving wave. Panel (b) shows the
frequencies of the driving (straight line) and driven (dotted line) waves.
The continuous AR frequency locking is seen starting $\protect\tau \approx
-3 $.}
\end{figure}

\begin{figure}
\includegraphics[width=9cm]{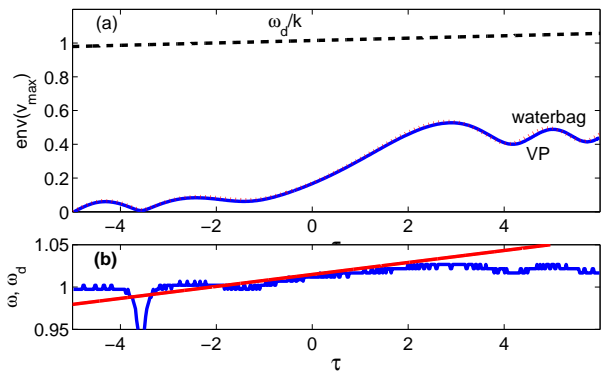}
\caption{(color online) The numerical simulations as in Fig. 3, but with the
driving amplitude below the AR threshold. Panel (a) illustrates saturation
of the excited wave amplitude, while panel (b) shows that the frequency
locking discontinues leading to saturation shortly after passage through the
linear resonance.}
\end{figure}
The evolution of the system leading to the final stage shown in Figs 1 and 2
is illustrated in Fig. 3, where panel (a) presents the envelope of the
maxima of fluid velocity $v$ during the evolution versus slow time. In the
same figure, we also show a similar envelope (dashed line) obtained by using
a simplified water bag model described in the next section. We observe an
excellent agreement between the two models even for very large excitations.
This agreement is important since the water bag model will be used below in
calculating the threshold for AR excitation of SPWs. Also shown in Fig. 3a
by dashed line is the driving wave phase velocity $u_{d}=\omega
_{d}/k=\omega _{d}$. One can see that at the final times of the excitation
the fluid velocity approaches the phase velocity, indicating again the
proximity to the Landau resonance. Finally, Fig. 3b shows the driving and
driven wave frequencies (the latter is calculated from the time differences
between successive peaks of the fluid velocity) versus $\tau $. It
illustrates the characteristic signature of autoresonant waves, i.e. the
frequency (phase) locking between the driven and driving waves, which starts
even prior passage trough the linear resonance $\omega _{d}=\omega _{0}$ (at
$\tau =0$) and continues to the fully nonlinear stage. We complete this
sections by Fig. 4, showing the results of the simulations similar to those
in Fig. 3 (the same parameters and initial conditions), but for a smaller
driving amplitude $\varepsilon =0.0011$. One can see in the figure that the
wave excitation saturates in this case (see panel a), as the phase locking
between the driven and driving wave discontinues (see panel b) shortly after
passage through the linear resonance. We find that the peak electron density
in this case reaches 1.7 of the initial density, as compared to 2.5 in the
AR case in Fig. 3. These results illustrate the existence of the
characteristic AR threshold $\varepsilon _{th}$ on the driving amplitude [$%
\varepsilon _{th}=0.0015$ for the parameters of the above simulations, see
Eq. (\ref{5.18}) below]. The autoresonance threshold is a weakly nonlinear
phenomenon studied in many applications \citep{Lazar125}. We have seen in
Figs. 3 and 4 that a simplified water bag model can be used in describing
the weakly nonlinear evolution of the driven SPWs instead of using the full
VP system. The next four sections will present the theory of weakly
nonlinear autoresonant SPWs based on this model.
\begin{figure}
\includegraphics[width=9cm]{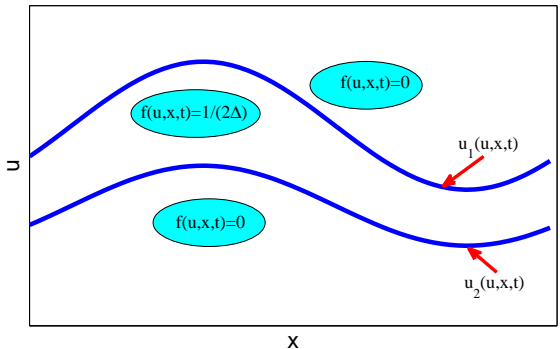}
\caption{(color online) The water bag model. The electron distribution is
confined between two limiting trajectories $u_{1,2}$.}
\label{WaterBagModel}
\end{figure}
\section{The water bag model of driven-chirped plasma waves}

The water bag model \citep{Berk} assumes that the electron distribution
function remains constant, $f(u,x,t)=1/(2\Delta )$, between two limiting
trajectories $u_{1,2}(x,t)$ in phase space and vanishes outside these
trajectories (see Fig. 5). Then the electron density is $n(x,t)=\int
fdu=(u_{1}-u_{2})/(2\Delta )$ and our kinetic SPW problem is governed by the
following set of the momentum and Poisson equations
\begin{eqnarray}
u_{1t}+u_{1}u_{1x}=(\varphi +\varphi _{d})_{x}, \nonumber \\
u_{2t}+u_{2}u_{2x}=(\varphi +\varphi _{d})_{x}, \\ 
\displaystyle \varphi _{xx}=\kappa ^{2}\varphi +(u_{1}-u_{2})/(2\Delta )-1, \nonumber
\end{eqnarray} 
\label{WB}
where we use the same normalization of $x,t,f,u_{1,2},\varphi ,$ and $\varphi
_{d}$ as in the previous section. We will be solving this system subject to $%
2\pi $ periodicity in $x$ for a trivial initial equilibrium $\varphi =0,$ $%
u_{1,2}=\pm \Delta $. Note that if one also defines $v(x,t)=(u_{1}+u_{2})/2$%
, Eqs. (\ref{WB}) yield
\begin{eqnarray}
n_{t}+(vn)_{x}=0,  \nonumber \\
v_{t}+vv_{x}=(\varphi +\varphi _{d})_{x}-\Delta ^{2}nn_{x},  \label{FluidM}
\\
\varphi _{xx}=\kappa ^{2}\varphi +n-1.  \nonumber
\end{eqnarray}%
Therefore, the water bag model is isomorphic to the usual warm fluid limit
of the driven plasma waves with the adiabatic electron pressure scaling $%
p\sim n^{3}$ and $\Delta ^{2}=3\sigma ^{2}$. The results shown by the dotted
red lines in Figs. 3 and 4 were obtained by solving Eqs. (\ref{WB})
numerically and show excellent agreement between the VP and water bag
simulations until approaching the Landau resonance. These water bag
simulations used a code similar to that for SIAWs \citep{Lazar151} and based
on a standard spectral method \citep{Canuto}.

In analyzing the AR\ threshold in the problem, we use a Lagrangian approach.
We introduce new potentials $\psi _{1,2}$ via $u_{1,2}=\pm \Delta +\partial
_{x}\psi _{1,2}$ and rewrite Eqs. (\ref{WB}) as
\begin{eqnarray}
\psi _{1tx}+\Delta \psi _{1xx}+\psi _{1x}\psi _{1xx}=(\varphi +\varphi
_{d})_{x},  \nonumber \\
\psi _{2tx}-\Delta \psi _{2xx}+\psi _{2x}\psi _{2xx}=(\varphi +\varphi
_{d})_{x},  \label{potentials} \\
\varphi _{xx}=\kappa ^{2}\varphi +(\psi _{1x}-\psi _{2x})/2\Delta .  \nonumber
\end{eqnarray}%
This system can be derived from the variation principle $\delta (\int
Ldxdt)=0$, with the three-field Lagrangian density
\begin{eqnarray}
L=\frac{1}{2}\varphi _{x}^{2}+\frac{\kappa ^{2}\varphi ^{2}}{2}-\frac{\psi
_{1x}^{2}+\psi _{2x}^{2}}{4}-\frac{\psi _{1x}\psi _{1t}-\psi _{2x}\psi _{2t}%
}{4\Delta }  \nonumber \\
-\frac{\psi _{1x}^{3}-\psi _{2x}^{3}}{12\Delta }+\frac{(\psi _{1x}-\psi
_{2x})(\varphi +\varphi _{d})}{2\Delta }.  \label{Lagr}
\end{eqnarray}%
Our next goal is to apply the Whitham's averaged variational principle %
\citep{Whitham} and derive weakly nonlinear slow evolution system describing
driven-chirped SPWs governed by this Lagrangian. We proceed by discussing
weakly nonlinear driven and phase-locked, but not chirped SPWs.

\section{Weakly nonlinear SPWs}

If one starts in the trivial equilibrium $\varphi =0,$ $n=1$, $v=0$ ($%
u_{1,2}=\pm \Delta $), Eqs. (\ref{FluidM}) after averaging over one spatial
period, yield constant-in-time averaged density and fluid velocity $%
\left\langle n\right\rangle =1$, $\left\langle v\right\rangle =0$ ($%
\left\langle u_{1,2}\right\rangle =\pm \Delta $). We consider the linear
stage of driven SPWs first, i.e., write $n=1+\delta n$ and linearize (\ref%
{FluidM}) to get
\begin{eqnarray}
(\delta n)_{t}+v_{x}=0,  \nonumber \\
v_{t}=\varphi _{x}-\Delta ^{2}\delta n_{x}-2\varepsilon \cos \theta _{d}\sin
x,  \label{4.1} \\
\varphi _{xx}=\kappa ^{2}\varphi +\delta n.  \nonumber
\end{eqnarray}%
In the case of a constant driving frequency, this set yields phase-locked
standing wave solutions of frequency $\omega =\omega _{d}$ for all dependent
variables%
\begin{eqnarray}
\delta n=a\cos \theta \cos x,  \nonumber \\
v=b\sin \theta \sin x,  \label{4.2} \\
\varphi =c\cos \theta \cos x,  \nonumber
\end{eqnarray}%
where $\theta =\theta _{d}=\omega t$,%
\begin{eqnarray}
a=-\frac{2\varepsilon }{\omega ^{2}-\omega _{0}^{2}},  \nonumber \\
b=\omega a,  \label{4.3} \\
c=-\frac{a}{1+\kappa ^{2}}.  \nonumber
\end{eqnarray}%
and%
\begin{equation}
\omega _{0}^{2}=\frac{1}{(1+\kappa ^{2})}+\Delta ^{2}.  \label{4.4}
\end{equation}%
Note that $\omega _{0}$ is the natural frequency of a linear SPW in the
problem. Equations (\ref{4.2}) yield the linear solutions
\begin{equation}
u_{1,2}=v\pm \Delta (1+\delta n)=\pm \Delta +b\sin \theta \sin x\pm \Delta
a\cos \theta \cos x  \label{4.5}
\end{equation}%
and, thus,%
\begin{equation}
\psi _{1,2}=-b\sin \theta \cos x\pm \Delta a\cos \theta \sin x.  \label{4.6}
\end{equation}%
Interestingly, in contrast to $n,u$, and $\varphi $, the linear solutions
for $u_{1,2}$ and $\psi _{1,2}$ are not standing waves. \qquad

Our next goal is to include a weak nonlinearity in the problem, but still
for a constant driving frequency case. To this end, we use spatial
periodicity and write truncated Fourier expansions

\begin{eqnarray}
\psi _{1,2}=b_{1,2}\cos x+a_{1,2}\sin x+d_{1,2}\cos (2x)+e_{1,2}\sin (2x),
\label{4.7} \\
\varphi =c_{1}\cos x+c_{2}\cos (2x),  \label{4.7a}
\end{eqnarray}%
where $a_{1,2}$, $b_{1,2}$, and $c_{1}$ are time dependent amplitudes viewed
as small first order objects, while $d_{1,2}$, $e_{1,2}$, and $c_{2}$ are
due to the nonlinearity and, thus, are of second order. The time dependence
of the first order amplitudes is assumed to be that of the linear solutions (%
\ref{4.6}) and (\ref{4.2}), i.e. $a_{1,2}=A_{1,2}\cos \theta $, $%
b_{1,2}=B_{1,2}\sin \theta $, and $c_{1}=C_{1}\cos \theta $. But what is the
time dependence of the second order amplitudes? Instead of working with the
original system (\ref{WB}) for answering this question, we use a simpler
Lagrangian approach in Appendix A. The final result is [see Eqs. (\ref{A3})]
\begin{eqnarray}
d_{1,2}=D_{1,2}\sin (2\theta )  \nonumber \\
e_{1,2}=F_{1,2}+E_{1,2}\cos (2\theta )  \label{4.8} \\
c_{2}=B+C\cos (2\theta )  \nonumber
\end{eqnarray}%
where $D_{1,2}$, $F_{1,2}$, $E_{1,2}$, $B$, and $C$ are constants. Thus, the
potentials characterizing a weakly nonlinear SPW phase-locked to the drive
are:
\begin{equation}
\psi _{i}=B_{i}\sin \theta \cos x+A_{i}\cos \theta \sin x+D_{i}\sin (2\theta
)\cos (2x)+[F_{i}+E_{i}\cos (2\theta )]\sin (2x),  \label{4.9a}
\end{equation}%
\begin{equation}
\varphi =C_{1}\cos \theta \cos x+[B+C\cos (2\theta )]\cos (2x).  \label{4.10}
\end{equation}%
Here $i=1,2$, all the coefficients (amplitudes) are constant and could be
related to the driving amplitude $\varepsilon $ by using the approach of
Appendix A. Nevertheless, we will not follow this route because, knowing the
form of Eqs. (\ref{4.9a}) and (\ref{4.10}) is sufficient for addressing our
original driven-chirped problem via the Whitham's averaged variational
principle \citep{Whitham}, as described next.

\section{Whitham's averaged variational principle for driven SPWs}

As in the case of the standing IAWs \citep{Lazar158},\ the Whitham's
approach uses the ansatz of form (\ref{4.9a}) and (\ref{4.10}) for the
solution of our driven-chirped problem, but now all the amplitudes are
viewed as \textit{slow} functions of time and $\theta $ is the\ \textit{fast}
wave phase having a slowly varying frequency $\omega (t)=d\theta /dt$
generally different from the frequency $\omega _{d}(t)$ of the chirped
driving wave. This ansatz is then substituted into the Lagrangian density (%
\ref{Lagr}), where the driver is written as $\varphi _{d}=2\varepsilon \cos
(\theta +\Phi )\sin x$ and the phase mismatch $\Phi (t)=\theta _{d}-\theta $
is viewed as a \textit{slow} function of time. Then, all the slow variables
are frozen in time and the Lagrangian density is averaged over $2\pi $ in
both $x$ and $\theta $ (fast scales). This results in a new averaged
Lagrangian density $\Lambda $, which depends on slow objects only [i.e., $%
A_{1,2}$, $B_{1,2}$, $C_{1}$ (first order amplitudes), $D_{1,2}$, $E_{1,2}$,
$F_{1,2}$, $B$, $C$ (second order amplitudes), the wave frequency $\omega $,
and the phase mismatch $\Phi $]. To fourth order in amplitudes we obtain
[via Mathematica \citep{Mathematica}] $\Lambda =\Lambda _{2}+\Lambda
_{4}+\Lambda _{d}$, where

\begin{eqnarray}
\Lambda _{2}=2C_{1}^{2}(1+\kappa
^{2})-A_{1}^{2}-A_{2}^{2}-B_{1}^{2}-B_{2}^{2}  \nonumber \\
-\frac{2}{\Delta }[\omega (A_{1}B_{1}-A_{2}B_{2})-C_{1}(A_{1}-A_{2})],
\label{5.1}
\end{eqnarray}%
\begin{eqnarray}
\Lambda
_{4}=-4(2F_{1}^{2}+2F_{2}^{2}+D_{1}^{2}+D_{2}^{2}+E_{1}^{2}+E_{2}^{2})
\nonumber \\
+2(4+\kappa ^{2})(2B^{2}+C^{2})-\frac{4}{\Delta }[2\omega
(D_{1}E_{1}-D_{2}E_{2})  \nonumber \\
+2B(F_{1}-F_{2})+C(E_{1}-E_{2})]-\frac{1}{\Delta }%
[A_{1}B_{1}D_{1}-A_{2}B_{2}D_{2}]  \nonumber \\
+\frac{1}{2\Delta }%
[B_{1}^{2}(2F_{1}-E_{1})-A_{1}^{2}(2F_{1}+E_{1})-B_{2}^{2}(2F_{2}-E_{2})+A_{2}^{2}(2F_{2}+E_{2})],
\label{5.1aa}
\end{eqnarray}%
\begin{equation}
\Lambda _{d}=\frac{4\varepsilon }{\Delta }(A_{1}-A_{2})\cos \Phi .
\label{5.1b}
\end{equation}%
Note that $\Lambda $ has an algebraic form with the wave phase $\theta $
entering via $\Phi =$ $\theta _{d}-\theta $ and $\omega (t)=d\theta /dt$.
Therefore, the variations with respect to all amplitudes yield a set of
eight algebraic equations
\begin{equation}
\frac{\partial \Lambda }{\partial Z_{i}}=0,i=1,..,8,  \label{5.2}
\end{equation}%
where $Z_{i}$ represent each of the $8$ slow amplitudes in the problem. In
addition, the variation with respect to $\theta $ yields an ordinary
differential equation (ODE)%
\begin{equation}
\frac{d}{dt}\left( \frac{\partial \Lambda }{\partial \omega }\right) =-\frac{%
\partial \Lambda }{\partial \Phi }.  \label{5.3}
\end{equation}%
To lowest order, the last equation becomes%
\begin{equation}
\frac{d}{dt}(A_{1}B_{1}-A_{2}B_{2})=-2\varepsilon (A_{1}-A_{2})\sin \Phi .
\label{5.4}
\end{equation}%
and, as expected, describes the \textit{slow} evolution, provided $%
\varepsilon $ is sufficiently small.

The plan for analyzing our slow driven system is as follows. First, we will
use the algebraic equations (\ref{5.2}) for expressing seven of the eight
amplitudes and $\omega $ in terms of $\Phi $ and the eighth amplitude
(chosen to be $C_{1}$), i.e. obtain relations $Z_{i}=Z_{i}(C_{1},\Phi )$ and
$\omega =\omega (C_{1},\Phi )$. Then%
\begin{equation}
\frac{d\Phi }{dt}=\omega _{d}(t)-\omega (C_{1},\Phi ),  \label{5.5}
\end{equation}%
which in combination with Eq. (\ref{5.4}) comprise a closed set of two ODEs
describing the slow evolution in our driven system. This plan is
algebraically complex, but can be performed using Mathematica %
\citep{Mathematica}. We describe the intermediate steps of this calculation
in Appendix B and here present the final results. To lowest significant
order, the amplitudes in Eq. (\ref{5.4}) are as in the linear problem [see
Eqs. (\ref{B2}), (\ref{B3})]%
\begin{eqnarray}
A_{1}=-A_{2}\approx -\frac{C_{1}\Delta }{\omega _{0}^{2}-\Delta ^{2}}%
=-C_{1}\Delta (1+\kappa ^{2}),  \label{5.6} \\
B_{1}=B_{2}\approx \frac{C_{1}\omega _{0}}{\omega _{0}^{2}-\Delta ^{2}}%
=C_{1}\omega _{0}(1+\kappa ^{2}),  \label{5.7}
\end{eqnarray}%
while [see Eqs. (\ref{B9}) and (\ref{B10}) in Appendix B]
\begin{equation}
\omega ^{2}(C_{1},\Phi )\approx \omega _{0}^{2}+NC_{1}^{2}+\frac{%
2\varepsilon \cos \Phi }{(1+\kappa ^{2})C_{1}},  \label{5.8}
\end{equation}%
where

\begin{equation}
N=\frac{9\kappa ^{2}(3+\kappa ^{2})+R}{96[1+(4+\kappa ^{2})\Delta ^{2}]},
\label{5.9}
\end{equation}%
and $R=(1+\kappa ^{2})\Delta ^{2}[3(60+56\kappa ^{2}+11\kappa
^{4})+20(1+\kappa ^{2})(4+\kappa ^{2})(5+2\kappa ^{2})\Delta
^{2}+16(1+\kappa ^{2})^{2}(4+\kappa ^{2})^{2}\Delta ^{4}]$.

\section{Dynamics and control of chirped-driven SPWs}

At this stage, we discuss the slow evolution in our driven-chirped SPWs
system. Upon substitution of Eqs. (\ref{5.6}) and (\ref{5.7})\ into Eq. (\ref%
{5.4}), we obtain%
\begin{equation}
\frac{dC_{1}}{dt}=-\frac{\varepsilon }{\omega _{0}(1+\kappa ^{2})}\sin \Phi .
\label{5.10}
\end{equation}%
Next, for a small frequency deviation in the vicinity of the linear
resonance, $\omega =\omega _{0}+\delta \omega $, Eq. (\ref{5.8}) yields%
\begin{equation}
\delta \omega \approx \frac{N}{2\omega _{0}}C_{1}^{2}+\frac{\varepsilon \cos
\Phi }{\omega _{0}(1+\kappa ^{2})C_{1}},  \label{5.11}
\end{equation}%
where the two terms on the right represent the wave frequency shifts due to
the nonlinearity and interaction, respectively. Then, from Eq. (\ref{5.5})
and using the driving frequency $\omega _{d}=\omega _{0}+\alpha t$, we obtain

\begin{equation}
\frac{d\Phi }{dt}=\alpha t-\frac{N}{2\omega _{0}}C_{1}^{2}-\frac{\varepsilon
\cos \Phi }{\omega _{0}(1+\kappa ^{2})C_{1}}.  \label{5.12}
\end{equation}%
\begin{figure}
\includegraphics[width=9cm]{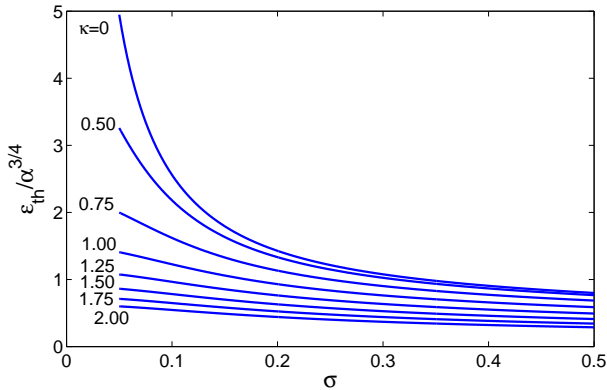}
\caption{(color online) The rescaled threshold driving amplitude $\protect%
\varepsilon _{th}/\protect\alpha ^{3/4}$ versus $\protect\sigma=k\protect
\lambda_{D}$ for different values of the screening parameter $\kappa$}
\end{figure}
Equations (\ref{5.10}) and (\ref{5.12}) comprise a complete set of
amplitude-phase mismatch ODEs describing the passage through the linear
resonance in our problem. These equations involve several parameters, but
the number of parameters can be reduced to just one by rescaling the
problem. Indeed, if we use the slow time $\tau =\alpha ^{1/2}t$ and define
new amplitude $Q$ via $Q^{2}=\frac{N}{2\omega _{0}\alpha ^{1/2}}C_{1}^{2}$,
our system reduces to
\begin{eqnarray}
\frac{dQ}{d\tau }=-\mu \sin \Phi ,  \label{5.14} \\
\frac{d\Phi }{d\tau }=\tau -Q^{2}-\frac{\mu \cos \Phi }{Q}.  \label{5.15}
\end{eqnarray}%
where
\begin{equation}
\mu =\frac{\varepsilon N^{1.2}}{\sqrt{2}\alpha ^{3/4}\omega
_{0}^{3/2}(1+\kappa ^{2})}.  \label{5.16}
\end{equation}%
Note that, if one defines a complex variable $\Psi =Qe^{i\Phi }$, our system
is further reduced to a single complex ODE%
\begin{equation}
i\Psi _{\tau }+(\tau -|\Psi |^{2})\Psi =\mu   \label{5.17}
\end{equation}%
characteristic to AR problems in many different physical systems and studied
in numerous application \citep{Lazar125}. For example, if initially $\Psi =0$
and one starts at sufficiently large negative $\tau $ (i.e. far from the
linear resonance), this equation predicts transition to AR at large positive
$\tau $ if $\mu $ is above the threshold $\mu _{th}\approx 0.41$, or
returning to our original parameters,
\begin{equation}
\varepsilon >\varepsilon _{th}=0.58\frac{\alpha ^{3/4}\omega
_{0}^{3/2}(1+\kappa ^{2})}{N^{1.2}}.  \label{5.18}
\end{equation}%
The threshold (\ref{5.18}) assumes its simplest form when either $\Delta $
or $\kappa $ vanish. Indeed, in the cold plasma case, $\Delta =0$ and
\begin{equation}
N=\frac{3\kappa ^{2}(3+\kappa ^{2})}{32}.  \label{5.19}
\end{equation}%
Therefore, for small $\kappa $, $\varepsilon _{th}$ scales as $\varepsilon
_{th}\sim \alpha ^{3/4}/\kappa $ and one needs a non-vanishing screening
factor $\kappa $ to get a sufficiently small $\varepsilon _{th}$ for the AR
excitation. In the case of $\kappa =0$ and $\Delta \ll 1$,%
\begin{equation}
N\approx \frac{15\Delta ^{2}}{8},  \label{5.20}
\end{equation}%
so $\varepsilon _{th}$ scales as $\varepsilon _{th}\sim \alpha ^{3/4}/\Delta
$ (recall that $\Delta =3^{1/2}k\lambda _{D}$ and measures the electron
thermal spread in the problem). We illustrate these results in Fig. 6
showing $\varepsilon _{th}/\alpha ^{3/4}$ vs $\sigma $ (this is $k\lambda _{D}$ in dimensional notations) for several values of $\kappa $.
\begin{figure}
\includegraphics[width=9cm]{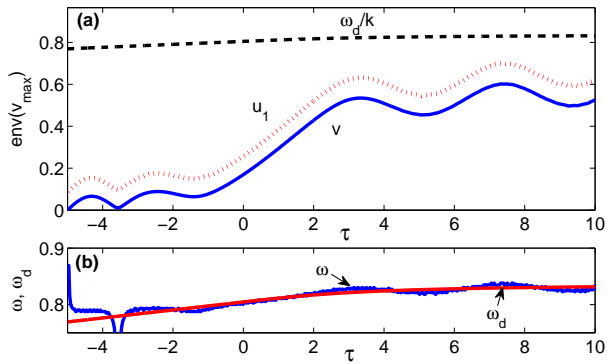}
\caption{(color online) The control of the autoresonant SPW by tapering the driving
frequency. Panel (a): The envelope of the maxima of
the electron fluid velocity $v$ versus slow time $\protect\tau =\protect%
\alpha ^{1/2}t$. The dashed line represents the phase velocity $\protect%
\omega _{d}/k$ of the driving wave and the dotted red line represents the
envelope of the maxima of the upper limiting velocity of the water bag
model. Panel (b) shows the frequencies of the driving (red line) and driven
(blue line) waves. The waves are frequency locked starting $\protect\tau %
\approx -3$.}
\end{figure}

Finally, we address the possibility of autoresonant control of large
amplitude SPWs. This control uses a one-to-one correspondence between the
amplitude and frequency of nonlinear waves. Since in autoresonance the
driven wave frequency is locked to the driving frequency, the wave amplitude
can be controlled by simply varying the driving frequency. For example, we
have seen previously (see Fig. 3), that a continuous \textit{linear} time
variation of the driving frequency leads to the excited wave approaching the
wave breaking limit. But what if one wants to avoid this limit and excite a
wave having some given target amplitude? This goal can be achieved by
tailoring the time variation of the driving frequency appropriately. We
illustrate such control in Fig. 7 showing the results of the simulations
using the water bag model for the case of $\sigma =0.05$, $\kappa =0.75$, $%
\varepsilon =0.0011$ ($\varepsilon _{th}=0.00084$ in this case) and the
driving frequency $\omega _{d}=\omega _{0}+\alpha t$ for $t<0$, but $\omega
_{d}=\omega _{0}+\frac{2d}{\pi }arctg(t/T)$ for $t>0$, where $T=$ $\frac{2d}{%
\pi \alpha }$. This driving frequency approaches a fixed value of $\omega
_{0}+d$ at large times (we used $d=0.033$ in the simulations). Fig. 7a shows
the envelopes of the maxima of the electron fluid velocity $v$ (solid line)
and of the upper limiting velocity $u_{1}$ of the water bag model (dotted
line). In the same figure, we also show the driving wave phase velocity
(dashed line). Separate simulations show that if the driving frequency in
this example is chirped linearly in time, the upper limiting velocity of the
water bag approaches the Landau resonance ($u_{1}\approx \omega _{d}/k$) at $%
\tau =6$, where wave breaking is expected. In contrast, for the saturated
driving frequency case shown in Fig. 7, the excited fluid velocity also
saturates at larger times and performs slow oscillations around some fixed
value. The frequency of these oscillations scales as $\varepsilon ^{1/2}$ %
\citep{Lazar125} and their presence illustrates stability of the AR
evolution. The velocity $u_{1}$ also saturates and, thus, the kinetic wave
breaking is avoided. This saturation is caused by the autoresonant frequency
locking in the system as illustrated in Fig. 7b, showing the evolution of
the frequencies of the driven and driving waves.

\section{Conclusions}

We have studied excitation and control of large amplitude SPWs by a chirped
frequency driving wave. The process involved passage through the linear
resonance in the problem and transition to autoresonant stage of excitation,
where the driven SPW self-adjusts its amplitude to stay in a continuous
resonance with the drive. The method allowed reaching extreme regimes, where
the electron density developed a sharply peaked spatial profile with the
maximum electron density exceeding the initial plasma density significantly
(see Fig. 1). These results were illustrated in both Vlasov-Poisson and
water bag simulations. The simpler water bag model [Eqs. (\ref{WB})] was
used for developing the adiabatic theory using the Whitham's averaged
variational principle for studying the weakly nonlinear stage of formation
of autoresonant SPWs. In this regime, the problem was reduced to the
standard set of coupled amplitude-phase mismatch equations (\ref{5.14}) and (%
\ref{5.15}) characteristic to many other autoresonantly driven problems. The
reduction also allowed finding the threshold driving amplitude [see Eq. (\ref%
{5.18})] for the transition to autoresonance. By slowly decreasing the chirp
rate of the driving frequency and reaching some fixed frequency level, one
could arrive at a given target amplitude of the autoresonant SPW and avoid
the kinetic wave breaking (see Fig. 7).

The self-consistent inclusion of the variation of the driving amplitude and
of 3D effects in the process of autoresonant excitation of SPWs seem to be
an important goals for future research. The form of the autoresonant SPW
suggests that it comprises a two-phase solution, with each phase locked to
one of the traveling waves comprising the drive. A better understanding of
such waveforms, analyzing other autoresonant multi-phase plasma waves, and
studying details of the kinetic wave breaking process in application to
autoresonant SPWs also comprise interesting goals for the future.

\begin{acknowledgements}
This work was supported by the US-Israel Binational Science Foundation grant No. 6079 and the Russian state program AAAA-A18-
118020190095-4. The authors are also grateful to J.S. Wurtele, P. Michel, and G. Marcus for helpful comments and suggestions.
\end{acknowledgements}

\begin{appendix}
\section{Second order amplitudes}\label{appA}

For calculating the second order amplitudes, we substitute Eqs. (\ref{4.7})
and (\ref{4.7a}) into the Lagrangian (\ref{Lagr}), write the result to
second nonlinear order in amplitudes and average over one spatial period.
This spatially averaged Lagrangian governs the \textit{time} dependence of
all the amplitudes. Its variations with respect to the second order
amplitudes yield a system of ODEs for these amplitudes versus time.
Algebraically, this reduction is a tedious process done here using
Mathematica \citep{Mathematica}. The resulting set of equations is%
\begin{eqnarray}
(e_{1,2})_{t}=\frac{1}{2}(a_{1,2}b_{1,2}\pm 4\Delta d_{1,2}),  \nonumber \\
(d_{1,2})_{t}=\frac{1}{4}(4c_{2}-a_{1,2}^{2}+b_{1,2}^{2}\mp 8\Delta e_{1,2}),
\label{A1} \\
c_{2}=\frac{e_{2}-e_{1}}{\Delta (4+\kappa ^{2})},  \nonumber
\end{eqnarray}%
or after substituting the first order dependencies $a_{1,2}\sim \cos \theta $%
, $b_{1,2}\sim \sin \theta $, and $c_{1}\sim \cos \theta $

\begin{eqnarray}
(e_{1,2})_{t}=p_{1,2}\sin (2\theta )\pm 2\Delta d_{1,2},  \nonumber \\
(d_{1,2})_{t}=q_{1,2}+r_{1,2}\cos (2\theta )+c_{2}\mp 2\Delta e_{1,2},
\label{A2} \\
c_{2}=\frac{e_{2}-e_{1}}{\Delta (4+\kappa ^{2})},  \nonumber
\end{eqnarray}%
where $p_{1,2}$, $q_{1,2}$, $r_{1,2}$ are constants. These equations have
the following time periodic solutions%
\begin{eqnarray}
d_{1,2}=D_{1,2}\sin (2\theta ),  \nonumber \\
e_{1,2}=F_{1,2}+E_{1,2}\cos (2\theta ),  \label{A3} \\
c_{2}=B+C\cos (2\theta ).  \nonumber
\end{eqnarray}%
where $D_{1,2}$, $F_{1,2}$, $E_{1,2}$, $B$, and $C$ are constants.

\section{The reduction of the variational system}\label{appB}

All algebraic manipulations in this Appendix use Mathematica %
\citep{Mathematica}. We employ the averaged Lagrangian $\Lambda $ [see Eqs. (%
\ref{5.1})-(\ref{5.1b})] and proceed from the variations with respect to the
first order amplitudes $A_{1,2}$, $B_{1,2}$, $C_{1}$:

\begin{eqnarray}
-2C_{1}+2A_{1}F_{1}+B_{1}D_{1}+A_{1}E_{1}+2B_{1}\omega +2A_{1}\Delta
=4\varepsilon \cos \Phi ,  \nonumber \\
-2C_{1}+2A_{2}F_{2}+B_{2}D_{2}+A_{2}E_{2}+2B_{2}\omega -2A_{2}\Delta
=4\varepsilon \cos \Phi ,  \nonumber \\
2F_{1}B_{1}-A_{1}(D_{1}+2\omega )-B_{1}(E_{1}+2\Delta )=0,  \label{B1} \\
-2F_{2}B_{2}+A_{2}(D_{2}+2\omega )+B_{2}(E_{2}-2\Delta )=0,  \nonumber \\
A_{1}-A_{2}+2C_{1}(1+\kappa ^{2})\Delta =0.  \nonumber
\end{eqnarray}%
The first four equations in this set yield the linear approximation in terms
of $C_{1}$%
\begin{eqnarray}
A_{1}^{\prime }=-A_{2}^{\prime }=-\frac{C_{1}\Delta }{\omega ^{2}-\Delta ^{2}%
},  \label{B2} \\
B_{1}^{\prime }=B_{2}^{\prime }=\frac{C_{1}\omega }{\omega ^{2}-\Delta ^{2}}.
\label{B3}
\end{eqnarray}%
Then the fifth equation in Eq. (\ref{B1}) gives the linear dispersion
relation%
\begin{equation}
\omega _{0}^{2}=\frac{1}{1+\kappa ^{2}}+\Delta ^{2}.  \label{B4}
\end{equation}%
Next, we take variations with respect to the second order amplitudes $%
D_{1,2} $, $E_{1,2}$, $F_{1,2}$, $B$, $C$ (to lowest significant order using
the linear result for the first order amplitudes and $\omega $) to get

\begin{eqnarray}
A_{1}^{\prime }B_{1}^{\prime }+8E_{1}\omega _{0}+8D_{1}\Delta =0,  \nonumber \\
A_{2}^{\prime }B_{2}^{\prime }+8E_{2}\omega _{0}-8D_{2}\Delta =0,  \nonumber \\
-8C0+A_{1}^{\prime 2}+B_{1}^{\prime 2}+16D_{1}\omega _{0}+16E_{1}\Delta =0,
\nonumber \\
-8C0+A_{2}^{\prime 2}+B_{2}^{\prime 2}+16D_{2}\omega _{0}-16E_{2}\Delta =0,
\label{B5} \\
8B-A_{1}^{\prime 2}+B_{1}^{\prime }-16F_{1}\Delta =0,  \nonumber \\
8B-A_{2}^{\prime 2}+B_{2}^{\prime 2}+16F_{2}\Delta =0,  \nonumber \\
F_{1}-F_{2}+B(4+\kappa ^{2})\Delta =0,  \nonumber \\
E_{1}-E_{2}+C(4+\kappa ^{2})\Delta =0,  \nonumber
\end{eqnarray}%
This system is now solved for the second order amplitudes via $C_{1}^{2}$:

\begin{eqnarray}
D_{1}=\frac{C_{1}^{2}\omega _{0}[2+\omega _{0}^{2}(4+\kappa ^{2})+3\Delta
^{2}(4+\kappa ^{2})]}{16(\omega _{0}^{2}-\Delta ^{2})^{2}[1-\omega
_{0}^{2}(4+\kappa ^{2})+\Delta ^{2}(4+\kappa ^{2})]},  \nonumber \\
E_{1}=-\frac{C_{1}^{2}\Delta \lbrack 3\omega _{0}^{2}(4+\kappa ^{2})+\Delta
^{2}(4+\kappa ^{2})]}{16(\omega _{0}^{2}-\Delta ^{2})^{2}[1-\omega
_{0}^{2}(4+\kappa ^{2})+\Delta ^{2}(4+\kappa ^{2})]},  \nonumber \\
F_{1}=\frac{C_{1}^{2}(4+\kappa ^{2})\Delta }{16(\omega _{0}^{2}-\Delta
^{2})[1+\Delta ^{2}(4+\kappa ^{2})]},  \label{B6} \\
D_{2}=D_{1}, E_{2}=-E_{1}, F_{2}=-F_{1},  \nonumber \\
B=-\frac{C_{1}^{2}}{8(\omega _{0}^{2}-\Delta ^{2})[1+\Delta ^{2}(4+\kappa
^{2})]},  \nonumber \\
C=\frac{C_{1}^{2}(3\omega _{0}^{2}+\Delta ^{2})}{8(\omega _{0}^{2}-\Delta
^{2})^{2}[1-\omega _{0}^{2}(4+\kappa ^{2})+-\Delta ^{2}(4+\kappa ^{2})]}.
\nonumber
\end{eqnarray}%
Finally, we return to the first two equations in (\ref{B1}) and solve these
equations for $A_{1,2}$ to higher (third) order in $C_{1}$

\begin{equation}
A_{1}^{\prime \prime }=-\frac{C_{1}\Delta }{\omega ^{2}-\Delta ^{2}}+\frac{%
C_{1}G_{1}-4\varepsilon \Delta (\omega _{0}^{2}-\Delta ^{2})\cos \Phi }{%
2(\omega _{0}^{2}-\Delta ^{2})^{2}},  \label{B7}
\end{equation}%
\begin{equation}
A_{2}^{\prime \prime }=\frac{C_{1}\Delta }{\omega ^{2}-\Delta ^{2}}+\frac{%
C_{1}G_{2}+4\varepsilon \Delta (\omega _{0}^{2}-\Delta ^{2})\cos \Phi }{%
2(\omega _{0}^{2}-\Delta ^{2})^{2}},  \label{B8}
\end{equation}%
where $G_{1}=\omega _{0}^{2}(2F_{1}-E_{1})+2D_{1}\omega _{0}\Delta -\Delta
^{2}(2F_{1}+E_{1})$ and $G_{2}=\omega _{0}^{2}(2F_{2}-E_{2})-2D_{2}\omega
_{0}\Delta -\Delta ^{2}(2F_{2}+E_{2})$. Finally, we use Eqs. (\ref{B6}) in
Eqs. (\ref{B7}) and (\ref{B8}) and substitute the resulting $A_{1,2}^{\prime
\prime }$ into the last equation in (\ref{B1}). This yields a higher order
approximation for the frequency $\omega $ of the wave:

\begin{equation}
\omega ^{2}=\omega _{0}^{2}+NC_{1}^{2}+\frac{2\varepsilon \cos \Phi }{%
(1+\kappa ^{2})C_{1}},  \label{B9}
\end{equation}%
where
\begin{equation}
N=\frac{9\kappa ^{2}(3+\kappa ^{2})+R}{96[1+(4+\kappa ^{2})\Delta ^{2}]},
\label{B10}
\end{equation}%
and $R=(1+\kappa ^{2})\Delta ^{2}[3(60+56\kappa ^{2}+11\kappa
^{4})+20(1+\kappa ^{2})(4+\kappa ^{2})(5+2\kappa ^{2})\Delta
^{2}+16(1+\kappa ^{2})^{2}(4+\kappa ^{2})^{2}\Delta ^{4}]$.
\end{appendix}

\bibliographystyle{jpp}

\bibliography{FSjpp}

\begin{thebibliography}{19}
\expandafter\ifx\csname natexlab\endcsname\relax\def\natexlab#1{#1}\fi
\def\au#1{#1} \def\ed#1{#1} \def\yr#1{#1}\def\at#1{#1}\def\jt#1{\textit{#1}}
  \def\bt#1{#1}\def\bvol#1{\textbf{#1}} \def\vol#1{#1} \def\pg#1{#1}
  \def\publ#1{#1}\def\arxiv#1{#1}\def\org#1{#1}\def\st#1{\textit{#1}}

\bibitem[Berk {\em et~al.\/}(1970)Berk, Nielsen \& Roberts]{Berk}
{\sc \au{Berk, H.}, \au{Nielsen, C.} \& \au{Roberts, K.}} \yr{1970}  \at{Phase
  space hydrodynamics of equivalent nonlinear systems: experimental and
  computational observations}.  \jt{Phys. Fluids}  \bvol{13},  \pg{980--995}.

\bibitem[Canuto {\em et~al.\/}(1988)Canuto, Hussaini, Quarteroni \&
  Zang]{Canuto}
{\sc \au{Canuto, C.}, \au{Hussaini, M.}, \au{Quarteroni, A.} \& \au{Zang, T.}}
  \yr{1988} {\em {Spectral Methods in Fluid Dynamics}\/}.
  \publ{Springer-Verlag}.

\bibitem[Dubin \& Ashourvan(2015)]{Dubin}
{\sc \au{Dubin, D.} \& \au{Ashourvan, A.}} \yr{2015}  \at{Trivelpiece-gould
  waves: frequency, functional form, and stability}.  \jt{Phys. Plasmas}
  \bvol{22},  \pg{102102}.

\bibitem[Fajans \& Friedland(2001)]{Lazar92}
{\sc \au{Fajans, J.} \& \au{Friedland, L.}} \yr{2001}  \at{Autoresonant
  (nonstationary) excitation of pendulums, plutinos, plasmas, and other
  nonlinear oscillators}.  \jt{Am. J. Phys.}  \bvol{69},  \pg{1096--1102}.

\bibitem[Friedland(2009)]{Lazar125}
{\sc \au{Friedland, L.}} \yr{2009}  \at{Autoresonance in nonlinear systems}.
  \jt{Scholarpedia}  \bvol{4},  \pg{5473}.

\bibitem[Friedland {\em et~al.\/}(2006)Friedland, Khain \& Shagalov]{Lazar111}
{\sc \au{Friedland, L.}, \au{Khain, P.} \& \au{Shagalov, A.}} \yr{2006}
  \at{Autoresonant phase-space holes in plasmas}.  \jt{Phys. Rev. Lett.}
  \bvol{96},  \pg{225001}.

\bibitem[Friedland {\em et~al.\/}(2019)Friedland, Marcus, Wurtele \&
  Michel]{Lazar158}
{\sc \au{Friedland, L.}, \au{Marcus, G.}, \au{Wurtele, J.} \& \au{Michel, P.}}
  \yr{2019}  \at{Excitation and control of large amplitude standing ion
  acoustic waves}.  \jt{Phys. Plasmas}  \bvol{26},  \pg{092109}.

\bibitem[Friedland \& Shagalov(2014)]{Lazar142}
{\sc \au{Friedland, L.} \& \au{Shagalov, A.}} \yr{2014}  \at{Excitation and
  control of chirped nonlinear ion-acoustic waves}.  \jt{Phys. Rev. E}
  \bvol{97},  \pg{063201}.

\bibitem[Friedland \& Shagalov(2017)]{Lazar151}
{\sc \au{Friedland, L.} \& \au{Shagalov, A.}} \yr{2017}  \at{Extreme driven ion
  acoustic wavess}.  \jt{Phys. Plasmas}  \bvol{24},  \pg{082106}.

\bibitem[Lehmann \& Spatachek(2016)]{Lehmann2016}
{\sc \au{Lehmann, G.} \& \au{Spatachek, K.}} \yr{2016}  \at{Transient plasma
  photonic crystals for high-power lasers}.  \jt{Phys. Rev. Lett.}  \bvol{116},
   \pg{225002}.

\bibitem[Lehmann \& Spatschek(2018)]{Lehmann2018}
{\sc \au{Lehmann, G.} \& \au{Spatschek, K.}} \yr{2018}  \at{Hamiltonian
  stochastic processes induced by succesive wave-particle interactions in
  stimulated raman scattering}.  \jt{Phys. Rev. E}  \bvol{79},  \pg{046404}.

\bibitem[Malkin {\em et~al.\/}(1999)Malkin, Shvets \& Fisch]{Malkin}
{\sc \au{Malkin, V.}, \au{Shvets, G.} \& \au{Fisch, N.}} \yr{1999}  \at{Fast
  compression of laser beams to highly overcritical power}.  \jt{Phys. Rev.
  Lett.}  \bvol{82},  \pg{4448--4451}.

\bibitem[Michel {\em et~al.\/}(2014)Michel, Divol, Turnbull \&
  Moody]{Michel2014}
{\sc \au{Michel, P.}, \au{Divol, L.}, \au{Turnbull, D.} \& \au{Moody, J.}}
  \yr{2014}  \at{Dynamic control of the polarization of intense laser beams via
  optical wave mixing in plasmas}.  \jt{Phys. Rev. Lett.}  \bvol{113},
  \pg{205001}.

\bibitem[Michel {\em et~al.\/}(2009)Michel, Divol, Williams, Weber, Thomas,
  Callahan, Haan, Salmonson, Dixit, Hinkel, Edwards, Macgowan, Lindl, Glenzer
  \& Suter]{Michel2009}
{\sc \au{Michel, P.}, \au{Divol, L.}, \au{Williams, E.}, \au{Weber, S.},
  \au{Thomas, C.}, \au{Callahan, D.}, \au{Haan, S.}, \au{Salmonson, J.},
  \au{Dixit, S.}, \au{Hinkel, D.}, \au{Edwards, M.}, \au{Macgowan, B.},
  \au{Lindl, J.}, \au{Glenzer, S.} \& \au{Suter, L.}} \yr{2009}  \at{Tuning the
  implosion symmetry of icf targets via controlled crossed-beam energy
  transfer}.  \jt{Phys. Rev. Lett.}  \bvol{102},  \pg{025004}.

\bibitem[Novikov {\em et~al.\/}(1984)Novikov, Manakov, Pitaevskii \&
  Zakharov]{Novikov}
{\sc \au{Novikov, S.}, \au{Manakov, S.}, \au{Pitaevskii, L.} \& \au{Zakharov,
  V.}} \yr{1984} {\em {Theory of solitons}\/}.  \publ{Plenum Publishing}.

\bibitem[Turnbull {\em et~al.\/}(2017)Turnbull, Goyon, Kemp, Pollock, Mariscal,
  Divol, Ross, Patankar, Moody \& Michel]{Turnbull2017}
{\sc \au{Turnbull, D.}, \au{Goyon, C.}, \au{Kemp, G.}, \au{Pollock, B.},
  \au{Mariscal, D.}, \au{Divol, L.}, \au{Ross, J.}, \au{Patankar, S.},
  \au{Moody, J.} \& \au{Michel, P.}} \yr{2017}  \at{Refractive index seen by a
  probe beam interacting with a laser-plasma system}.  \jt{Phys. Rev. Lett.}
  \bvol{118},  \pg{015001}.

\bibitem[Turnbull {\em et~al.\/}(2016)Turnbull, Michel, Chapman, Tubman,
  Pollock, Chen, Goyon, Ross, Divol, Woolsey \& Moody]{Turnbull2016}
{\sc \au{Turnbull, D.}, \au{Michel, P.}, \au{Chapman, T.}, \au{Tubman, E.},
  \au{Pollock, B.}, \au{Chen, C.}, \au{Goyon, C.}, \au{Ross, J.}, \au{Divol,
  L.}, \au{Woolsey, N.} \& \au{Moody, J.}} \yr{2016}  \at{High power dynamic
  polarization control using plasma photonics}.  \jt{Phys. Rev. Lett.}
  \bvol{116},  \pg{205001}.

\bibitem[Whitham(1974)]{Whitham}
{\sc \au{Whitham, G.}} \yr{1974} {\em {Linear and Nonlinear Waves}\/}.
  \publ{John Wiley \& Sons}.

\bibitem[{Wolfram Research Inc.}(2017)]{Mathematica}
{\sc \au{{Wolfram Research Inc.}}} \yr{2017}  \at{Mathematica}.  \jt{Version
  11.1} .

\end{thebibliography}

\end{document}